# Active control of excitonic strong coupling and electroluminescence in electrically driven plasmonic nanocavities

*Junsheng Zheng[1], Ruoxue Yang[1], Alexey V. Krasavin[2], Zhenxin Wang[1], Yuanjia Feng[1], Longhua Tang[1], Linjun Li[1,3], Xin Guo[1,3], Daoxin Dai[1], Anatoly V. Zayats[2,\*], Limin Tong[1,4,\*] and Pan Wang[1,3,4,\*]*

[1]Interdisciplinary Center for Quantum Information, New Cornerstone Science Laboratory, State Key Laboratory of Extreme Photonics and Instrumentation, College of Optical Science and Engineering, Zhejiang University, Hangzhou 310027, China

[2]Department of Physics and London Centre for Nanotechnology, King's College London, Strand, London WC2R 2LS, UK

[3]Jiaxing Key Laboratory of Photonic Sensing & Intelligent Imaging, Intelligent Optics & Photonics Research Center, Jiaxing Research Institute Zhejiang University, Jiaxing 314000, China

[4]Collaborative Innovation Center of Extreme Optics, Shanxi University, Taiyuan 030006, China

*Corresponding authors

E-mail: a.zayats@kcl.ac.uk; phytong@zju.edu.cn; nanopan@zju.edu.cn

**Abstract**

Enhancement and active control of light-matter interactions at the atomic scale is important for developing next-generation nanophotonic and quantum optical devices. Here, we demonstrate electric control of both excitonic strong coupling and electroluminescence by integrating semiconductor monolayers into a nanometer gap of electrically driven nanocube-on-mirror plasmonic nanocavities. Particularly, in a strongly-coupled system of nanocavity plasmons and $WSe_2$ excitons, the ultra-strong electric field generated in the nanocavity gap enables a reversible modulation of the Rabi splitting between ~102 and 80 meV with a bias below 2.5 V. In the quantum tunnelling regime, by injecting carriers into a nanocavity-integrated $WS_2$ monolayer, bias-controlled spectrally tunable electroluminescence from charged or neutral excitons is achieved with an external quantum efficiency reaching ~3.5%. These results underline practical approaches to electric control of atomic-scale light-matter interactions for applications including nanoscale light sources, ultrafast electro-optic modulation, quantum information processing and sensing.

The study of interactions between light and matter plays a pivotal role in the advance of optical physics and technologies. Squeezing optical fields down to a deep subwavelength-scale with plasmonic or dielectric nanostructures enables access to the extreme light-matter interactions at the atomic scale[1-7], which is of great interest for both state-of-the-art fundamental research and development of high-performance nanophotonic and quantum optical devices ranging from nanoscale light sources and optical modulators to photodetectors. In this context, atomic-scale light-matter interactions between solid-state emitters and a photonic nanostructure with high local density of optical states has been recently attracting considerable interest[5,6]. In such a coupled system, in addition to the weak coupling regime where the emitters exhibit Purcell-enhanced spontaneous emission[8-10], strong coupling can be achieved to form hybridized light-matter states when the rate of energy exchange between light and matter exceeds the dissipation rates in the system[11-14]. At the same time, for further fundamental exploration and practical applications, it is of immense importance to add another degree of freedom to actively control atomic-scale light-matter interactions represented by strong coupling and spontaneous emission, especially in an electric way, which is promising to uncover new physical phenomena via their complex interactions and lead to a variety of applications in nanoscale optoelectronics and quantum optical technologies.

In contrast to dielectric photonic nanostructures, metallic plasmonic nanostructures can perform simultaneous optical and electric functions, which have long been envisioned to merge photonics and electronics at the nanoscale within the same architecture[15,16], and opens up vast opportunities for robust electric control of atomic-scale light-matter interactions. Among various plasmonic nanostructures, plasmonic nanocavities, produced by two metallic objects separated by a nanometer-thick dielectric gap, such as nanoparticle-on-mirror nanocavities and metallic nanoparticle dimers, have been attracting increasing research interest recently due to their ability to provide extremely confined optical fields in the gap for enhancing light-matter interactions at the atomic scale[17,18] and to manifest fascinating quantum mechanical effects[19-21]. Moreover, upon application of a bias voltage of just several volts across the nanometer-thick gap, plasmonic nanocavities offer an attractive means to achieve ultra-strong electric fields up to the level of V/nm, additionally enabling engineered electron tunnelling. This provides unprecedented opportunities for the electric control of atomic-scale light-matter interactions originated from the strong interplay between ultra-confined optical fields, ultra-strong electric fields and various functional materials in the ultra-confined regions, which is difficult to be achieved with other systems.

Here, by integrating semiconductor monolayers into the nanometer gap of electrically driven single-crystal nanocube-on-mirror (NCoM) plasmonic nanocavities, we demonstrate electric control of both excitonic strong coupling and electroluminescence with high robustness. Particularly, upon functionalization of the nanocavity gap with a $WSe_2$ monolayer, strong coupling between nanocavity plasmons and $WSe_2$ excitons is achieved. More importantly, using an ultra-strong electric field (~0.5 V/m) produced in the gap with a bias of 2.5 V, the Rabi splitting can be reversibly modulated between ~102 and 81 meV by the control of the number of involved excitons. In the quantum tunnelling regime, electric excitation of excitonic luminescence from a nanocavity-integrated $WS_2$ monolayer with a Purcell-enhanced external quantum efficiency (EQE) as high as 3.5% is demonstrated by injecting carriers into the monolayer. Its spectral characteristics have been engineered via selective generation of charged and neutral excitons by controlling the bias polarity.

**Results**

**Fabrication of electrically driven plasmonic nanocavities.** Figure 1a shows a schematic illustration of an electrically driven plasmonic nanocavity. The NCoM plasmonic nanocavity is formed by depositing a gold nanocube, which is capped with a bilayer of cetyltrimethylammonium bromide (CTAB), on the top of a gold mirror covered by an alumina spacer and, if required, with a layer of functional material (e.g., a semiconductor monolayer in this work). For electric integration, the gold nanocube is partially embedded in an insulating poly(methyl methacrylate) (PMMA) layer and then coated with a transparent conductive layer of indium tin oxide (ITO), acting as a top electrode. When a bias is applied between the top and bottom electrodes, a strong electric field is generated in the nanocavity gap, while at the same time, the nanocavity can still be optically accessed through the transparent ITO layer to produce extremely confined electromagnetic fields in the same gap.

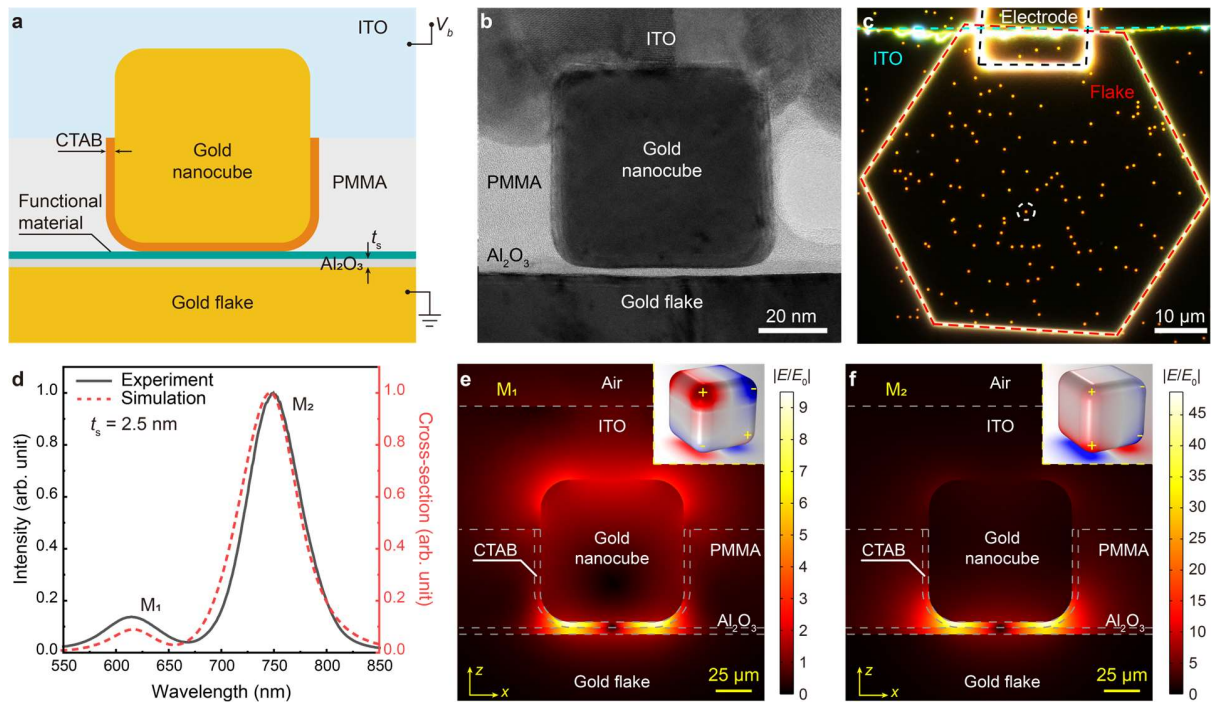

**Figure 1 | Fabrication and characterization of electrically driven plasmonic nanocavities. a**, Schematic illustration of an electrically driven plasmonic nanocavity, formed by a NCoM structure (if necessary, integrated with a functional material) contacted with a transparent conductive layer of ITO at the top, simultaneously providing optical and electric access. **b**, Cross-sectional TEM image of an electrically driven plasmonic nanocavity with an alumina layer thickness ($t_s$) of 2.5 nm. **c**, Dark-field scattering image of an ensemble of electrically driven plasmonic nanocavities ($t_s$ = 2.5 nm)

formed on a 50-nm-thick gold flake. The black, red and blue dashed lines indicate the outlines of the gold electrode, gold flake and ITO layer, respectively. **d**, Experimentally measured (solid line) and numerically calculated (dashed line) scattering spectra of the electrically driven plasmonic nanocavity encircled in **c**. **e,f**, Numerically simulated normalized near-field distribution of the electric field in the *xz* plane corresponding to nanocavity modes $M_1$ (**e**) and $M_2$ (**f**) marked in **d**. Inset, the corresponding normalized charge density distribution on the surfaces of the gold nanocube and flake.

In practice, it is very challenging to fabricate electrically driven plasmonic nanocavities by using deposited metal films (having a granular polycrystalline structure) as the mirror. The latter not only introduce a significant optical loss that can degrade the strength of light-matter interactions[22,23], but also make the nanocavity easily damaged or broken by surface roughness-induced protrusions in the nanocavity gap under a strong electric field. In this work, chemically synthesized gold nanocubes (with an average size of 60 nm and a CTAB thickness of ~1.8 nm) and gold flakes[7,22], having both a single-crystal structure and an ultrasmooth surface, were used as building blocks for the formation of electrically driven plasmonic nanocavities with low optical loss and high-quality nanoscale electric junctions. The procedure for the fabrication of electrically driven plasmonic nanocavities is explained in Methods. A cross-sectional transmission electron microscopy (TEM) image of an electrically driven plasmonic nanocavity (see Methods and Fig. 1b) clearly shows that the gold nanocube is perfectly separated by a dielectric layer (including 1.8-nm-thick CTAB and 2.5-nm-thick alumina layers) from the gold flake forming a uniform nanocavity gap with ultrasmooth interfaces, and is also electrically contacted with the top ITO layer. Benefitting from high transparency of the ITO layer, the nanophotonic modes of the electrically driven plasmonic nanocavities can be optically excited and observed as yellow spots in a dark-field scattering image (Fig. 1c, see Methods and Supplementary Fig. S1 for characterization details). The measured scattering spectrum (Fig. 1d, solid line) reveals two main scattering peaks located at 615 and 751 nm (labeled as modes $M_1$ and $M_2$, respectively), which match very well with the optical modes in numerically simulated scattering spectrum (Fig. 1d, dashed line, see Methods and Supplementary Fig. S2,3 for simulation details) and correspond to the excitation of flake-coupled quadrupolar (as indicated by the electric near-field distribution of mode $M_1$ shown in Fig. 1e) and transversal dipolar (as indicated by the electric near-field distribution of mode $M_2$ shown in Fig. 1f) modes of the nanocube[22,23]. The simulations show that greatly enhanced local field can be achieved in the nanometer-thick dielectric gap, which is important for the enhancement of light-matter interactions at the atomic scale. By controlling the thickness of the dielectric gap, the resonance peaks of these plasmonic modes can be precisely tuned to desired wavelengths (Supplementary Fig. S4).

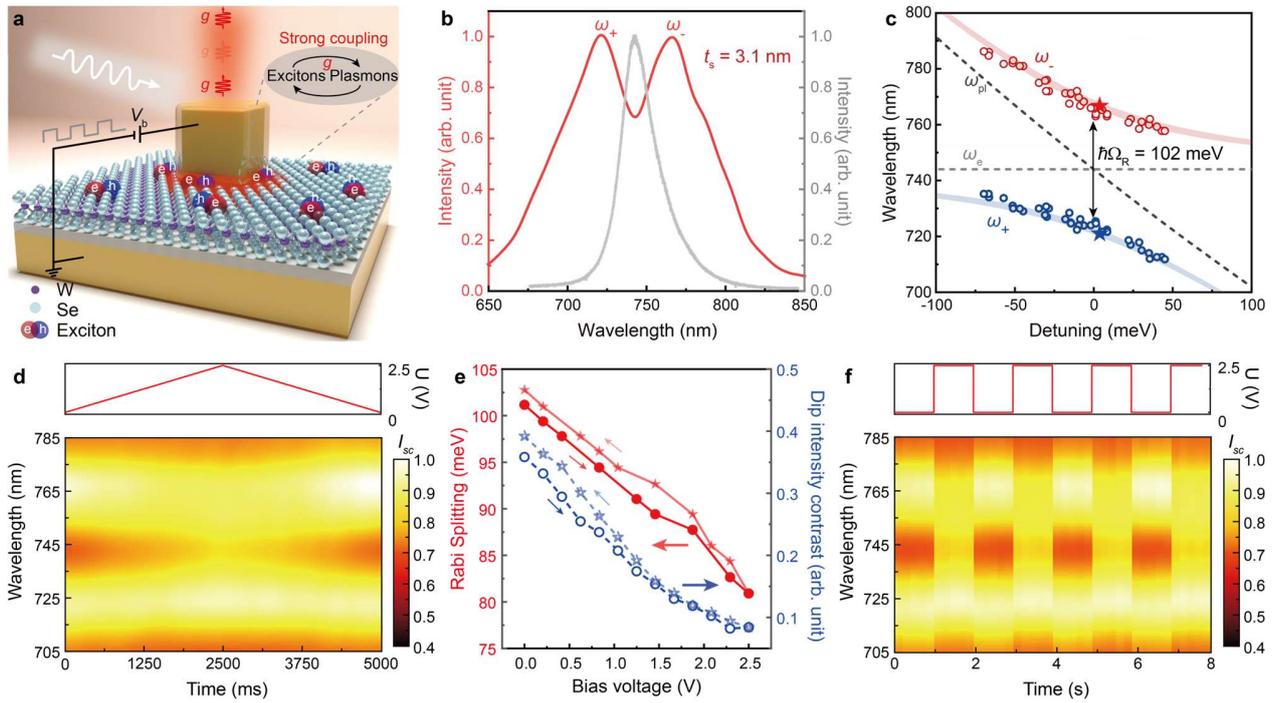

**Figure 2 | Electric modulation of strong plasmon-exciton coupling. a**, Schematic illustration of electric modulation of strong coupling between nanocavity plasmons and WSe$_2$ excitons. **b**, Measured scattering spectrum of an electrically driven plasmonic nanocavity, which is functionalized with a WSe$_2$ monolayer in the gap ($t_s$ = 3.1 nm, red solid line). For

reference, photoluminescence spectrum of a WSe$_2$ monolayer (gray solid line) is also provided. **c**, Measured spectral dispersion (hollow circles) of hybridized plasmon-exciton states as a function of the detuning between energies of the plasmonic mode (black dashed line) and excitons (grey dashed line). The colored lines represent analytically calculated spectral dispersion of the hybridized modes with Rabi splitting of ~102 meV. The solid stars represent the energies of two split peaks in **b**. **d**, Spectral map showing the evolution of the scattering spectra of the electrically driven plasmonic nanocavity (the same nanocavity as in **b**) with a continuously increasing and then a reversely decreasing bias as indicated in the upper panel. The color represents the normalized scattering intensity. **e**, Rabi splitting (red solid circles and stars) and dip intensity contrast (blue hollow circles and stars) extracted from the scattering spectra with forward and backward (indicated by downward and upward arrows, respectively) scanning of the bias during the modulation process presented in **d**. **f**, Spectral map showing the evolution of the scattering spectra with alternative switching of the bias between 0 and 2.5 V (as indicated in the upper panel). The color represents the normalized scattering intensity.

**Active modulation of strong plasmon-exciton coupling by ultra-strong electric field.** In addition to the extreme light confinement in the nanocavity[17,18], it is very straightforward to achieve an ultra-strong electric field inside the nanocavity gap upon application of a bias between the nanocube (using the top ITO electrode) and the gold flake, which opens opportunities for electric tuning of the properties of functional materials integrated in the gap and subsequently for active modulation of atomic-scale light-matter interactions critical for many applications. Here, we exploit these advantages for electric modulation of strong plasmon-exciton coupling, which is highly required for practical applications of strong coupling in optoelectronics and quantum information processing[24,25], but challenging to realize due to the limited tunability of excitons in functional materials with other approaches.

For the experimental demonstration, a monolayer of WSe$_2$, which hosts in-plane oriented A excitons at room temperature (with an emission peak around 1.67 eV) due to strong Coulomb interaction and reduced dielectric screening[26], was integrated into the nanometer gap of the electrically driven plasmonic nanocavities, as schematically illustrated in Fig. 2a. On the other hand, the spectral position of the plasmonic resonances can be tuned by employing WSe$_2$-functionalized nanocavities with different alumina thickness. When the resonance peak of mode M$_2$ is matched to the emission peak of the A excitons in the WSe$_2$ monolayer (grey solid line in Fig. 2b), a typical mode splitting profile arising from strong coupling can be clearly observed in the nanocavity scattering spectrum (red solid line in Fig. 2b), presenting two hybridized plasmon-exciton modes. Figure 2c further presents the evolution of the peak wavelengths of the two plasmon-exciton branches (upper branch $\omega_-$ and lower branch $\omega_+$) with controlled detuning ($\delta = \omega_{pl} - \omega_e$) between the plasmon and exciton energies (the former were scanned by employing WSe$_2$-functionalized nanocavities with various combinations of the alumina thickness and nanocube size), which exhibits a distinct anti-crossing behavior with Rabi splitting $\hbar\Omega_R = 2\hbar g = 102$ meV (where $g$ represents the coupling strength). This value clearly satisfies the criteria for strong coupling $\Omega_R \geq |(\kappa + \gamma)/2|$[27], where the excitonic linewidth ($\hbar\kappa$) is ~57 meV and the average plasmon linewidth ($\hbar\gamma$) for mode M$_2$ is ~100 meV. This result is further confirmed by employing simplified Jaynes-Cummings model[28] to describe the strong coupling effect, which can reproduce the experimental results very well (red and blue lines in Fig. 2c).

Upon application of a bias between the top ITO layer and bottom gold flake, electric modulation of the strong plasmon-exciton coupling was further realized (Fig. 2d). With a gradual increase of the bias from 0 to 2.5 V (corresponding to the nanocavity electric field of ~0.5 V/nm), the Rabi splitting in the scattering spectrum gradually decreases, which is accompanied by an obvious decrease of the intensity contrast of the scattering dip (defined as $2(I_{\text{peak}} - I_{\text{dip}})/(I_{\text{peak}} + I_{\text{dip}})$, where $I_{\text{peak}}$ and $I_{\text{dip}}$ are the peak and dip intensity values in the normalized scattering spectra). When the applied bias is lifted, the strong plasmon-exciton coupling in the WSe$_2$-functionalized nanocavity recovers back to its initial state. The modulation of the Rabi splitting (between ~102 and 81 meV) and the corresponding dip intensity contrast (between ~0.36 and 0.08) with tuning of the bias can be seen more vividly in Fig. 2e, also showing excellent reproducibility of the effect. From a theoretical point of view, the Rabi splitting $\Omega_R$ is proportional to the square root of the number of excitons ($\sqrt{N}$) involved in the strong coupling and inversely proportional to the square root of the volume ($\sqrt{V}$) of the plasmonic mode, as $\Omega_R = \mu_m\sqrt{4\hbar N c/(\lambda \varepsilon \varepsilon_0 V)}$ (where $\lambda$ and $\mu_m$ are the transition wavelength and dipole moment of the excitons, respectively, $\varepsilon$ is the permittivity of the material surrounding the semiconductor monolayer, $\hbar$ is the reduced Planck constant, and $\varepsilon_0$ is the free-space permittivity)[12]. Therefore, in the case of an unchanged mode volume, the electric modulation of the strong plasmon-exciton coupling in the WSe$_2$-functionalized nanocavity is mainly due to the variation of the number of the involved excitons, which can be estimated to decrease from ~90 to 56 with the increase of bias from 0 to 2.5 V (see Methods and Supplementary Fig. S5). It is known that the stability of excitons can be greatly reduced with the application of a strong static electric field in the out-of-plane direction, which eventually leads to their dissociation and a decrease in the density of excitons in the WSe$_2$ monolayer[29,30]. This is experimentally confirmed by the gradual decrease in the photoluminescence intensity of a nanocavity-integrated WSe$_2$ monolayer under the application of an increasing bias (Supplementary Fig. S6). Figure 2f further presents a map showing the evolution of the scattering spectra from the electrically driven nanocavity, with the bias alternatively switched between 0 and 2.5 V, which illustrates the excellent repeatability in the electric modulation of the strongly coupled system.

Compared with existing active modulation approaches, such as electrostatic gating[31-33], thermal tuning[32,34] and electrochemical switching[35], the electric field-based modulation approach demonstrated here provides advantages including a much lower operation bias of just 2.5 V (facilitating the integration with nanoelectronic circuits), a higher modulation bandwidth (theoretically limited only by the RC time constant of the device, therefore potentially reaching the THz level) and an all-solid-state architecture. Moreover, different from the electrostatic gating approach that only works for conductive excitonic materials with at least micrometer-scale lateral size (necessary for electric contact)[31-33], the electric field-based modulation approach can be readily applied to strong coupling systems using excitonic materials such as molecules or quantum dots. Therefore, it represents a universal approach for strong coupling modulation, which is attractive not only for electric control of optical signals at the nanoscale, but also for applications such as quantum manipulation and quantum information processing.

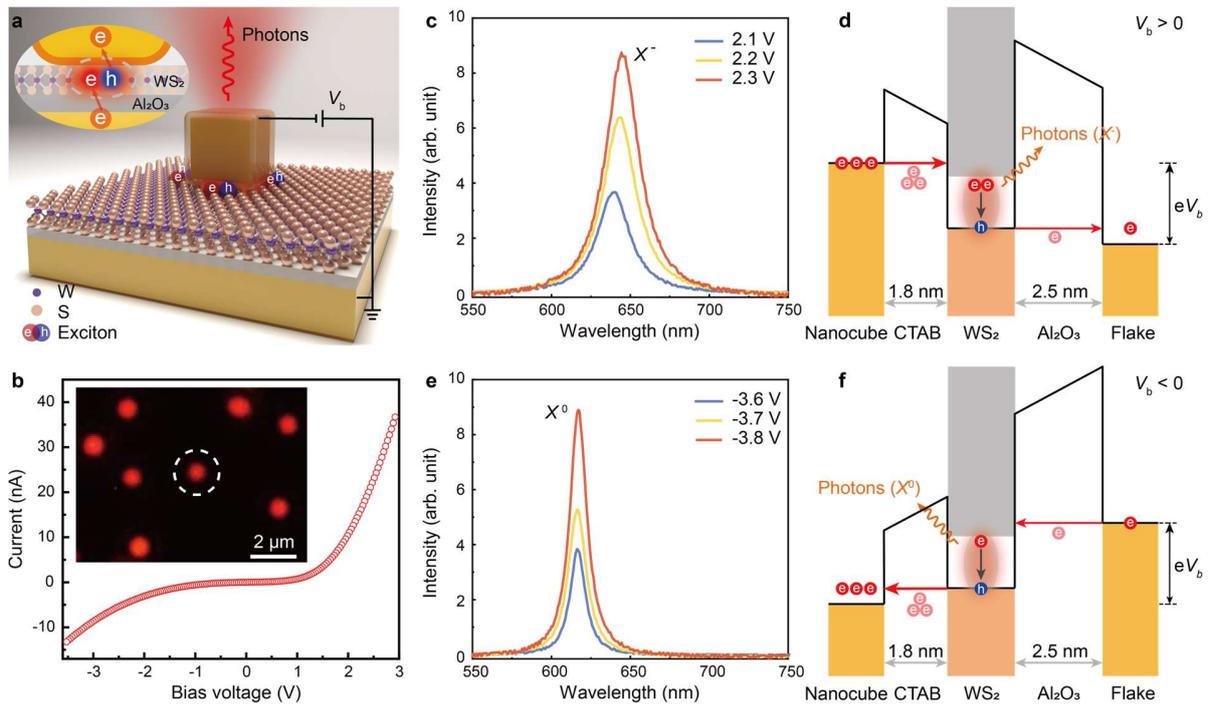

**Figure 3 | Electric excitation and control of excitonic luminescence by quantum tunnelling. a**, Schematic illustration of a quantum tunnelling-enabled excitonic luminescence process in an electrically driven plasmonic nanocavity

functionalized with a WS$_2$ monolayer. **b**, Current-voltage characteristic of a tunnelling device containing ~50 plasmonic nanocavities functionalized with a WS$_2$ monolayer. Inset, detected red light emission (captured with a colour CCD camera) from nanocavities under a forward bias of 2.3 V. **c,e**, Measured emission spectra from the nanocavity marked in **b** under various forward (**c**) and backward (**e**) biases. **d,f**, Energy diagrams of the electrically driven plasmonic nanocavity under a forward (**d**) and a backward (**f**) bias.

**Spectral tunable excitonic electroluminescence enabled by quantum tunnelling.** In the case of an ultrathin nanocavity gap, the ultra-strong electric field further enables quantum tunnelling, which can be exploited for the direct electric excitation of luminescence from nanocavity-integrated active materials via the injection of electrons and holes. Moreover, the plasmonic nanocavities featuring large Purcell factors provide an unprecedented ability for the engineering and enhancement of electroluminescence. Therefore, the architecture developed here also presents great interest for the realization of electrically driven nanoscale light sources with high quantum efficiency.

To demonstrate this, a monolayer of WS$_2$, which is a two-dimensional n-type semiconductor[36] (its photoluminescence is dominated by negatively charged excitons, as shown in Supplementary Fig. S7), was integrated into electrically driven plasmonic nanocavities having an alumina spacer with a decreased thickness ($t_s$ = 2.5 nm). Its excitonic luminescence can be excited by quantum tunnelling and then emitted to the optical far field via the excitation of nanocavity plasmonic modes, as shown in Fig. 3a. The characteristic current-voltage curve for a tunnelling device containing ~50 nanocavities is presented in Fig. 3b. With the increase of the bias voltage, the current shows a superlinear increase, which confirms the tunnelling nature of the electronic transport through the gap of the plasmonic nanocavities. Under an application of a forward bias of 2.3 V between the top and the bottom electrodes, spots of red-colour light emission (inset of Fig. 3b, see Methods and Supplementary Fig. S8 for the experimental setup), each corresponding to the position of an individual electrically driven nanocavity, were clearly observed. Figure 3c further presents the emission spectra of the nanocavity encircled in Fig. 3b for various applied forward biases, which exhibit an emission peak around 643 nm and a linewidth of ~25 nm. The observed behavior agrees well with the photoluminescence characteristics of $WS_2$ monolayers at room temperature (Supplementary Fig. S7), indicating that the electroluminescence originates from the radiative recombination of charged ($X^-$) excitons in the integrated $WS_2$ monolayer, which are created by the quantum tunnelling[37-40]. Particularly, as schematically presented in Fig. 3d, under a sufficiently high forward bias, electrons of the gold nanocube tunnel into the conduction band of the integrated $WS_2$ monolayer, while at the same time, electrons in the valance band of the $WS_2$ monolayer tunnel to the gold flake, generating holes in their original places, thus allowing the formation of excitons in the $WS_2$ monolayer. Due to the slight difference in the thicknesses of the CTAB (~1.8 nm) and alumina (~2.5 nm) layers, electrons in the conduction band of the n-type $WS_2$ monolayer accumulate more efficiently than holes in the valence band, resulting in the generation of negatively-charged excitons that emit light with a wavelength around 643 nm at room temperature[41,42].

Light emission was also observed from the same nanocavity when a backward bias was applied. However, the emission spectra showed a blue-shifted peak wavelength of ~617 nm and a narrower linewidth of ~13 nm (Fig. 3e), which is totally different from the emission characteristic of the charged excitons in $WS_2$ monolayers and can be assigned to the radiative recombination of their neutral ($X^0$) A-type counterparts[36,41]. As schematically shown in Fig. 3f, in contrast to the forward-biased case, holes in the valence band of the n-type $WS_2$ monolayer accumulate more efficiently than electrons in the conduction band under a backward bias, which results in compensation of the natural n-doping of $WS_2$ and generation of $X^0$ excitons that emit light around 617 nm observed in the experiment. Therefore, the difference in the thicknesses of the dielectric barriers surrounding the WS$_2$ monolayer plays an important role in the active control of the electroluminescence, which is difficult to realize with other electric excitation approaches[43,44]. The EQE (the ratio between emitted photons and tunnelled electrons) of the electrically driven nanocavity under a forward bias of 2.3 V is estimated to be ~3.5% (see Methods for details). It is much higher than EQEs of tunnelling-excited electroluminescence of semiconductor monolayers demonstrated previously[45-47] due to the Purcell enhancement provided by plasmonic mode M$_1$ of the low-loss single-crystal nanocavity. The EQE can be further improved by optimizing the structural parameters of the nanocavities (e.g., the thicknesses of the alumina, PMMA and ITO layers) to have a better match between the excitonic emission spectrum and the resonant wavelength of the plasmonic mode (Supplementary Fig. S9).

Electroluminescence has been also observed from nanocavities functionalized with other semiconductor monolayers, such as WSe$_2$, which shows an excitonic emission peak at 754 nm with a linewidth of ~59 nm (Supplementary Fig. S10), agreeing well with the photoluminescence characteristics of negatively-charged excitons[48]. The EQE under a forward bias of 3.2 V is estimated to be ~7.5%, which is higher than that for WS$_2$ monolayers demonstrated above due to a better match of the excitonic emission wavelength with the spectral position of the dipolar plasmonic mode M$_2$. It is worth noting that light emission can also be obtained from electrically driven plasmonic nanocavities without any functionalization in the gap via an inelastic electron tunnelling process[49], during which electrons lose part of their energy to excite nanocavity plasmonic modes, as shown in Supplementary Fig. S11a,b. As the resonant modes of the plasmonic nanocavities are spectrally quite broad, for the inelastic tunnelling-induced emission it is challenging to obtain narrow linewidths[50-54]. For example, the emission spectra measured from electrically driven plasmonic nanocavities with $t_s$ = 2.5 nm has a broad linewidth of ~93 nm (Supplementary Fig. S11c). In contrast, by functionalizing the tunnel junctions with a semiconductor monolayer, an about 7 times narrower linewidth related to the excitonic luminescence can be obtained (e.g., Fig. 3e). Furthermore, the EQE of the quantum tunnelling-excited excitonic luminescence is about 6 times of that of the inelastic tunnelling-induced light emission (which is around 0.6%). The spectral tunable excitonic luminescence from semiconductor monolayers facilitated by tunnelling is promising not only for the highly-compact electric excitation of plasmonic modes, but also for the realization of electrically-driven nanoscale light sources highly required for integrated electronic/photonic systems.

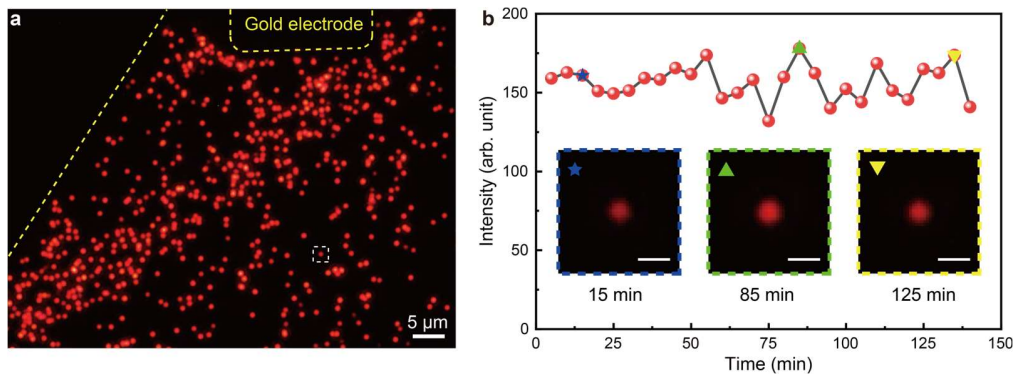

**Figure 4 | High robustness of electrically driven plasmonic nanocavities. a**, Image of light emission from hundreds of $WS_2$-functionalized plasmonic nanocavities excited by quantum tunnelling under a forward bias of 2.3 V. The yellow dashed lines indicate the outlines of the gold electrode and the flake. **b**, Time evolution of the detected integral light emission intensity from the plasmonic nanocavity marked in **a** under a continuously applied forward bias of 2.3 V. Insets, images of light emission from the nanocavity taken at 15, 85 and 125 minutes.

**Robustness of electrically driven plasmonic nanocavities.** The robustness of electrically driven plasmonic nanocavities is critical for their use in the active control of atomic-scale light-matter interactions and practical device applications. In general, it is challenging to fabricate nanoscale electric junctions with a high yield and operation stability due to the facile breakdown of nanometer gaps under an ultra-strong electric field, caused by surface roughness of metallic interfaces and nonuniformity in the thickness of the dielectric spacer. The high crystal quality and ultrasmooth interfaces, together with the precise alumina deposition approach, enables rigorous fabrication of nanoscale electric structures with a high yield and robustness. Taking the electroluminescence as an example, hundreds of $WS_2$-functionalized plasmonic nanocavities formed on a gold flake (over an area of ~2100 $\mu m^2$) can be simultaneously lighten up (seen as red emission spots) by quantum tunnelling with uniform distribution of the emission intensities across the ensemble (Fig. 4a), clearly indicating the excellent fabrication yield reaching almost 100%, which is important for practical applications. Furthermore, benefitting from the excellent structural quality, the electrically driven nanocavities can continuously work for long time without a visible decrease in the emission intensity, as demonstrated by 2.5-hour measurements presented in Fig. 4b, which indicates their excellent long-term working stability and robustness important for practical applications.

**Discussion**

Employing electric integration of single-crystal plasmonic nanocavities to simultaneously obtain ultra-confined optical fields and ultra-strong electric fields in the nanometer gap, we have demonstrated active control of atomic-scale light-matter interactions including electric modulation of strong plasmon-exciton coupling and spectral tunable excitonic electroluminescence enabled by quantum tunnelling. This approach can be generalized to electric control of other types of light-matter interactions such as nonlinear optical response. In additional semiconductor monolayers, other functional materials, such as dye molecules, quantum dots and rare earth ions, can also be readily integrated into the nanocavity gap.

Reciprocally, ultrafast optical control and generation of electric signals is also possible with electrically driven plasmonic nanocavities[52,55,56], which is of interest for applications such as ultrafast photodetection and lightwave electronics. In conclusion, electrically driven plasmonic nanocavities provide a versatile platform for strong enhancement and active control of atomic-scale light-matter interactions and open opportunities for developing next-generation optoelectronic and quantum optical devices that can seamlessly interface electronics and photonics at the nanoscale.

**Methods**

**Fabrication of electrically driven plasmonic nanocavities.** Firstly, a single-crystal gold flake was transferred onto a silicon substrate (covered with a 300-nm-thick $SiO_2$ layer) to work as the gold mirror and bottom electrode for electrically driven plasmonic nanocavities. Secondly, a gold strip electrode was fabricated at the edge of the gold flake for electric connection. Thirdly, a nanometer-thick alumina layer was deposited on the surface of the gold flake using an atomic layer deposition approach to work as a dielectric spacer. At this stage, if necessary, functional materials such as semiconductor monolayers can be integrated into the nanocavities by using a polydimethylsiloxane (PDMS)-assisted transfer method. Fourthly, a diluted solution of gold nanocubes was drop-casted onto the gold flake covered with an alumina layer (and, if needed, a functional material layer) to define the plasmonic nanocavities. Then, an insulating PMMA layer was spin-coated onto the sample, which was subsequently partially etched by $O_2$ plasma to expose the top of the gold nanocubes. Finally, a 50-nm-thick ITO layer (working as the top electrode) was sputtered on the top through a soft PDMS shadow mask, used to define its shape and position.

**TEM characterization.** For cross-sectional TEM characterization of electrically driven plasmonic nanocavities, an electron-transparent cross-sectional lamella of a selected nanocavity was prepared as follows. Firstly, electrically driven plasmonic nanocavities were fabricated on a silicon substrate using the approach introduced above. Secondly, a layer of carbon (~150 nm in thickness) was sputtered on the sample to protect the nanocavities. Then, an electrically driven plasmonic nanocavity was selected under SEM, and a cross-sectional lamella of the nanocavity with a thickness of ~300 nm was obtained using a focused-ion-beam (FIB) system (Helios G4, Thermo Scientific). Finally, the lamella was transferred onto a copper grid, and imaged using a high-resolution transmission electron microscope (F200X G2, Talos) operated at 200 kV.

**Numerical simulation of scattering spectra.** Numerical simulations of scattering spectra from electrically driven plasmonic nanocavities were performed using a finite element method (COMSOL Multiphysics software) in a scattered field formulation. The geometry of the electrically driven plasmonic nanocavity used for the simulations, schematically shown in Supplementary Fig. S2a, was set to match that of the experimentally structures (derived from cross-sectional TEM images). The nanocavity was illuminated with a TE-polarized electromagnetic plane wave at 80° to match the experimental conditions. The wavelength of the incident wave was varied from 500 to 900 nm, while the power flow of the scattered fields was integrated inside a 64° collection angle corresponding to the NA of the objective employed in the experiments, 500 nm from the center of the nanocube bottom face. The normalized electric near-field distributions ($|E/E_0|$) and the surface charge density distributions ($\rho$) for various modes were obtained from the full vectorial field given by the numerical solution. The refractive index of the single-crystal gold was taken from Olmon *et. al.*[57] and the refractive index of ITO was taken from the experimentally measured data (as shown in Supplementary Fig. S3). The refractive indices of silica, alumina, CTAB and PMMA were set to be 1.45, 1.70, 1.44, and 1.48, respectively.

**Estimation of number of excitons involved in strong coupling.** For the system which has $N$ excitons coupled with a single plasmonic mode, the total Rabi splitting can be approximately calculated as[12]:

$$\Omega_R = \mu_m \sqrt{\frac{4\pi \hbar N c}{\lambda \varepsilon \varepsilon_0 V}}, \tag{1}$$

where $\lambda$ and $\mu_m = 7.675\ D$ are the transition wavelength and the transition dipole moment of excitons in a WSe$_2$ monolayer[58], respectively, $\varepsilon$ is the permittivity of the medium surrounding the excitonic material, $\hbar$ is the reduced Planck constant, and $\varepsilon_0$ is the free-space permittivity. The effective mode volume of the plasmonic mode $V$ can be estimated by integrating the energy density $W(\mathbf{r})$ over the region of the nanogap first, and then normalizing the result to its maximum value[59]:

$$V = \frac{\int W(\mathbf{r}) d\mathbf{r}}{\max[W(\mathbf{r})]}, \tag{2}$$

Thus, using the experimental Rabi splitting and the calculated mode volume, the number of involved excitons during the modulation process can be calculated.

**Light emission characterization setup.** Upon the application of a static bias voltage from a source meter (2611B, Keithley), the light emission from individual nanocavities was collected by a 100× objective (NA = 0.9, TU Plan Fluor, Nikon) and directed through a beam splitter to a CCD camera for imaging and to a spectrometer (combination of a monochromator (Kymera 193i, Andor) and an electron multiplying charge coupled device (EMCCD) camera (iXon Ultra, Andor)) for spectral analysis, as schematically shown in Supplementary Fig. S8a. A color CCD camera (DS-Fi3, Nikon) was used for imaging of excitonic luminescence from a semiconductor monolayer, while an EMCCD camera (iXon Ultra, Andor) was used for the measurement of inelastic tunnelling-induced light emission. Then, the emission spectrum $P(\lambda)$ was obtained by correcting the measured emission spectrum $P_{\text{meas}}(\lambda)$ with a spectral response of the measurement setup, $T(\lambda)$, which includes the wavelength-dependent transfer efficiency of all optical elements in microscope and detection efficiency of the spectrometer. The obtained transfer function $T(\lambda)$ is presented in Supplementary Fig. S8b.

**Estimation of EQE for excitonic luminescence.** In the case of excitonic luminescence from an electrically interfaced plasmonic nanocavity functionalized with a WS$_2$ monolayer, the EQE can be calculated as a ratio between the number of the emitted photons and surface plasmon polaritons (SPPs) propagating along interface between gold flake and PMMA, $N_{\text{opt}}$, and the number of the tunnelled electrons, $N_e$, in a given period of time:

$$\text{EQE} = N_{opt}/N_e, \qquad (3)$$

To estimate $N_{\text{opt}}$, the light emission power from individual nanocavities, $P_{\text{meas}}$, was first measured employing an objective with NA = 0.9 and then corrected to obtain the emitted power in all directions, $P_{\text{opt}}$, using the transfer function, $T(\lambda_0)$ (where $\lambda_0$ is the wavelength of the emitted photons, Supplementary Fig. S8b) and the simulated angular emission characteristics of the plasmonic mode (see below). The number of the emitted photons and plasmons, $N_{\text{opt}}$, was calculated by dividing the corrected total emitted power, $P_{\text{opt}}$, by the photon energy (for simplicity, all of the emitted photons were assumed to have the same wavelength of $\lambda_0$ = 620 nm for the case of the forward bias (Fig. 3c,d) and $\lambda_0$ = 640 nm for the case of the backward bias (Fig. 3e,f)). The number of the tunnelled electrons can be extracted from the tunnelling current dividing it by the electron charge.

The numerical simulations of the angular emission characteristics were performed using a finite element method (COMSOL Multiphysics software) using a model schematically shown in Supplementary Fig. S2b. A dipole source was put in the middle of the dielectric gap to mimic the exciton emission source in the semiconductor monolayer, its wavelength was set be 620 nm for neutral excitons and 640 nm for charged excitons. In experiment, excitons can be generated everywhere in the semiconductor layer. Therefore, the ratio, $\eta_{opt}$, between emission power flow, $P_{\text{rad}}$, integrated inside a 64° collection angle corresponding to the NA of the objective and total power flow, $P_{\text{tot}}$, integrated over the upper semi-sphere 500-nm away from the bottom surface of the cube ($\eta_{opt} = P_{\text{rad}}/P_{\text{tot}}$) was obtained by modelling various dipole positions uniformly scanning the nanogap area (square grid, step 5 nm), and then the results were averaged to obtain the average ratio, $\eta_{opt\_ave}$. The corrected total emitted power, $P_{\text{opt}}$, can be obtained by dividing the measured emitted power, $P_{\text{meas}}$, by the transfer function, $T(\lambda_0)$, and the simulated average ratio, $\eta_{opt\_ave}$.

structures. *Small* **12**, 5190-5199 (2016).

**Acknowledgements** We thank Yeonhee Lee from the Department of Chemistry, Seoul National University for assistance on synthesis of gold nanocubes; Xi Mu from Westlake Center for Micro/Nano Fabrication for the facility support and technical assistance on sputtering and Yin Sun from Zhejiang University Center for Micro/Nano Fabrication for the technical assistance on laser-writing. This work was supported by the National Key Research and Development Project of China (Grant 2023YFB2806700), National Natural Science Foundation of China (Grants 62075195, 92250305 and 92150302), the UK EPSRC CPLAS project EP/W017075/1, New Cornerstone Science Foundation (NCI202216), Zhejiang Provincial Natural Science Foundation of China (Grant LDT23F04015F05). The data access statement: all the data supporting this research are presented in full in the results section and supplementary materials.

# Supplementary Information for
# Active control of excitonic strong coupling and electroluminescence in electrically driven plasmonic nanocavities

*Junsheng Zheng[1], Ruoxue Yang[1], Alexey V. Krasavin[2], Zhenxin Wang[1], Yuanjia Feng[1], Longhua Tang[1], Linjun Li[1,3], Xin Guo[1,3], Daoxin Dai[1], Anatoly V. Zayats[2,\*], Limin Tong[1,4,\*] and Pan Wang[1,3,4,\*]*

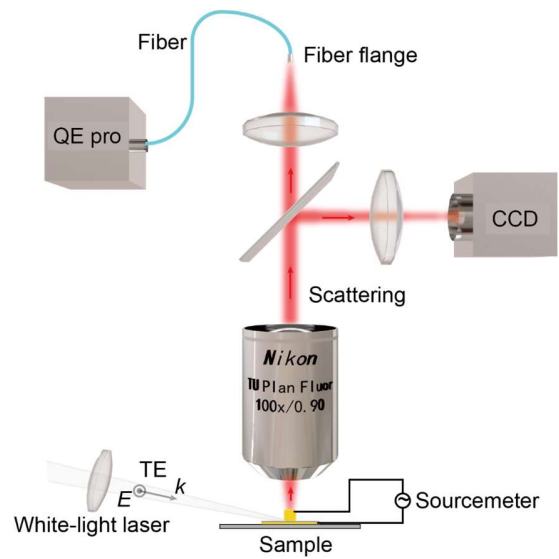

**Supplementary Fig. 1 | Dark-field spectroscopy setup.** Schematic diagram of the dark-field setup for light scattering imaging and spectroscopy of electrically driven plasmonic nanocavities. Briefly, a TE-polarized beam from a white-light laser (FIU-6, NKT Photonics) was first focused onto nanocavities at an oblique angle of about 80° with respect to the surface normal. The scattering signal from individual nanocavities was first collected by a 100× objective (NA = 0.9, TU Plan Fluor, Nikon), and then directed with a beam splitter to a charge-coupled device (CCD) camera (DS-Fi3, Nikon) for imaging and a spectrometer (QE Pro, Ocean Optics) for spectral analysis. All the measured scattering spectra were calibrated by the spectrum of the white-light laser and spectral response of the detection system.

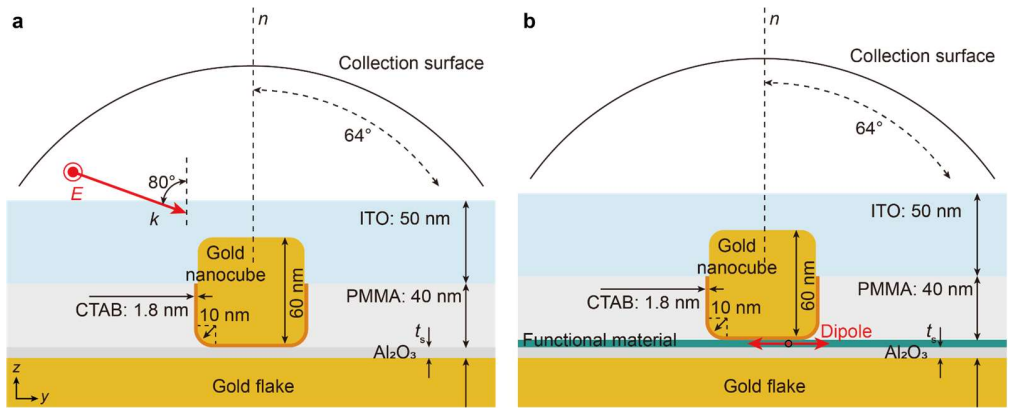

**Supplementary Fig. 2 | Numerical simulation models of plasmonic nanocavities. a,b**, Schematic illustrations of models used for (**a**) numerical simulations of scattering spectra, electric near-field and surface charge density distributions and (**b**) excitonic luminescence from a WS$_2$ monolayer.

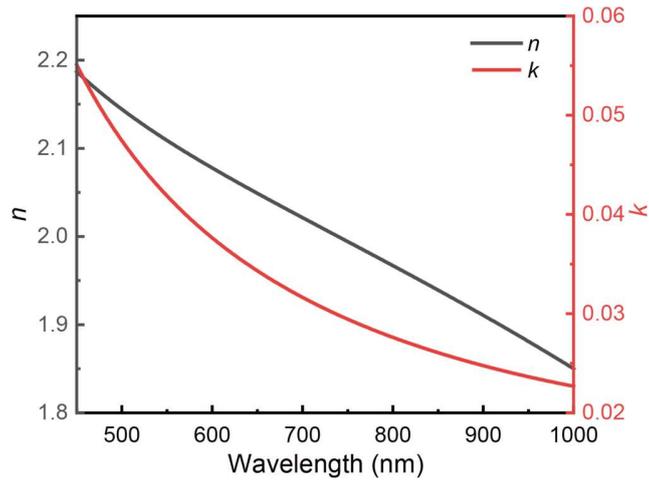

**Supplementary Fig. 3 | Refractive index of ITO electrode.** Measured real (black line) and imaginary (red line) parts of the refractive index of a 50-nm-thick ITO layer deposited by sputtering in a mixed gas atmosphere (98 sccm Ar and 2 sccm $O_2$) with DC power of 350 W for 180 s at room temperature.

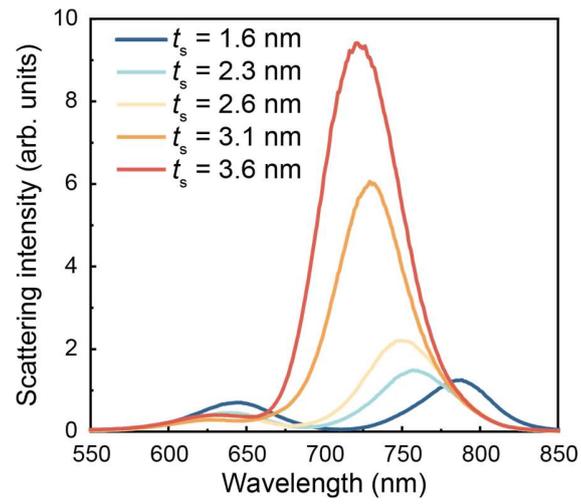

**Supplementary Fig. 4 | Effect of alumina thickness on nanocavity optical properties.** Experimentally measured scattering spectra of electrically driven plasmonic nanocavities with varying thickness of the alumina spacer.

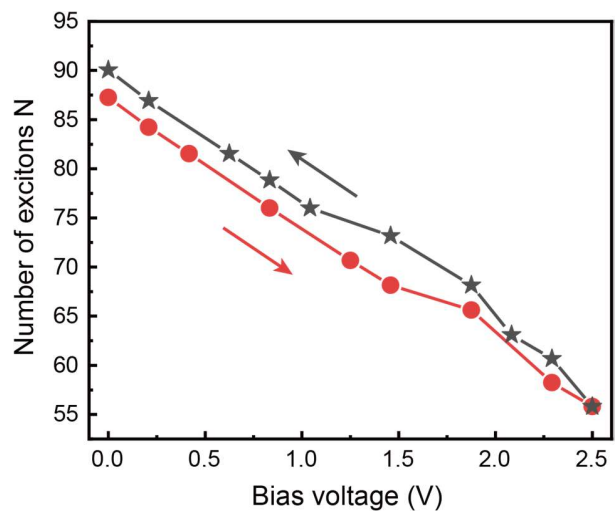

**Supplementary Fig. 5 | Number of excitons involved in modulation of strong coupling.** Estimated number of excitons involved in the modulation of strong coupling effect presented in Fig. 2d, which decreases from ~87 to ~56 with the increase of the bias voltage from 0 to 2.5 V, and then increases back to ~90 with the decrease of the bias voltage back to 0 V.

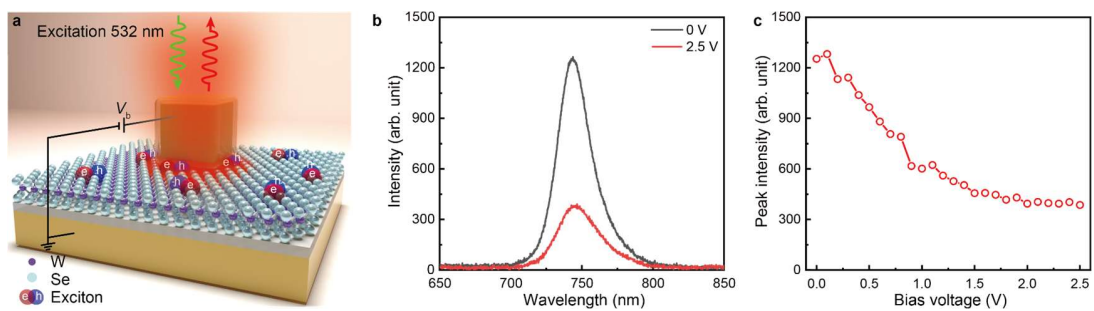

**Supplementary Fig. 6 | Electric modulation of photoluminescence in a WSe$_2$ monolayer. a**, Schematic illustration of the structure used for the investigation of electric modulation of photoluminescence in a WSe$_2$ monolayer. **b**, Photoluminescence spectra of a WSe$_2$ monolayer integrated in an electrically driven plasmonic nanocavity measured under bias voltages of 0 (black line) and 2.5 V (red line). **c**, Bias voltage-dependent peak intensities extracted from photoluminescence spectra of the WSe$_2$ monolayer.

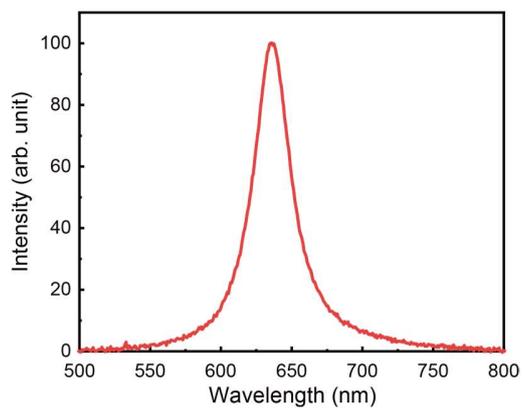

**Supplementary Fig. 7 | Photoluminescence spectrum of a WS$_2$ monolayer.** Photoluminescence spectrum of a WS$_2$ monolayer under a 532 nm excitation.

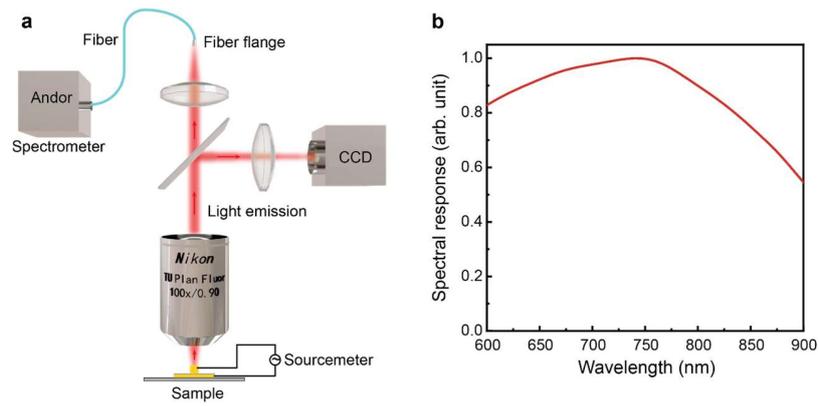

**Supplementary Fig. 8 | Light emission characterization setup. a**, Schematic diagram of the setup for imaging and spectroscopy of light emission from electrically driven plasmonic nanocavities. **b**, Spectral response of the measurement setup, which includes the spectral response of all optical elements in the microscope and detection efficiency of the spectrometer.

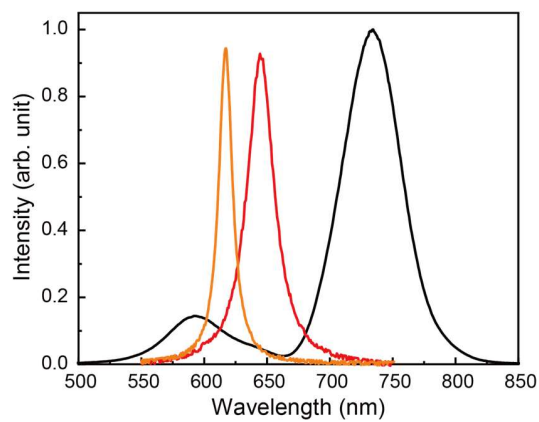

**Supplementary Fig. 9 | Scattering and electroluminescence spectra of electrically driven plasmonic nanocavities functionalized with a WS₂ monolayer.** Normalized scattering spectrum (black line) of the plasmonic nanocavity functionalized with a monolayer WS$_2$ (marked in the inset of Fig. 3b). The emission spectra from the nanocavity under 2.3 V (red line) and -3.8 V (orange line) biases are shown for comparison (the same curves from Fig. 3c,e).

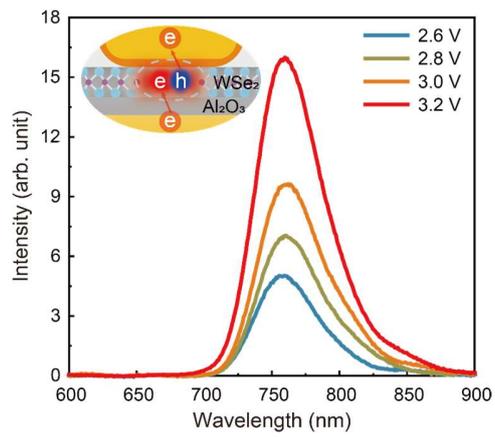

**Supplementary Fig. 10 | Electroluminescence spectra from a WSe₂ monolayer.** Bias-dependent excitonic electroluminescence spectra measured from an electrically driven plasmonic nanocavity functionalized with a WSe$_2$ monolayer (schematically shown in the inset).

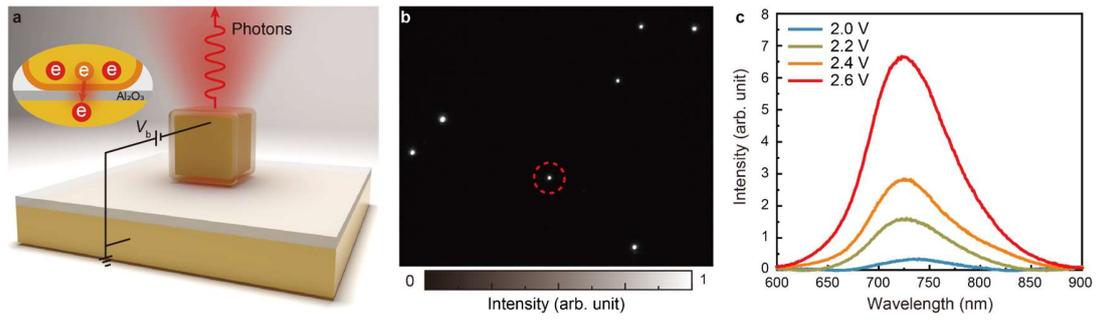

**Supplementary Fig. 11 | Inelastic tunnelling-induced light emission. a**, Schematic illustration of inelastic electron tunnelling-induced plasmon excitation and a light emission process in an electrically driven plasmonic nanocavities without a functional material. **b**, Image of inelastic electron tunnelling-induced light emission from the plasmonic nanocavities obtained by an electron multiplying charge coupled device (EMCCD) camera under a bias of 2.6 V. **c,** Measured emission spectra from a nanocavity marked in **b** under an increasing forward bias (from 2.0 to 2.6 V).